\documentclass[a4paper]{article}

\usepackage[pages=all, color=black, position={current page.south}, placement=bottom, scale=1, opacity=1, vshift=5mm]{background}
\SetBgContents{
	
}      

\usepackage[margin=1in]{geometry} 

\usepackage{amsmath}
\usepackage{amsthm}
\usepackage{amssymb}
\usepackage{hyperref}
\usepackage{longtable}
\usepackage{booktabs}
\usepackage{lscape}

\usepackage[utf8]{inputenc}
\usepackage{hyperref}
\hypersetup{
	unicode,
	pdfauthor={Ypma, Maaskant, Gill, Sjerps, van den Berg}
	pdfsubject={LRs for presence of body fluids from mRNA assay data in mixtures},
	pdfkeywords={Body fluid typing, mRNA profile, LR system, Machine learning, Calibration},
	pdfproducer={LaTeX},
	pdfcreator={pdflatex}
}

\usepackage[sort&compress,numbers,square]{natbib}
\theoremstyle{plain}

\theoremstyle{definition}

\usepackage{graphicx, color}
\graphicspath{{fig/}}

\usepackage{algorithm, algpseudocode} 
\usepackage{mathrsfs} 

\usepackage{lipsum}

\title{Calculating LRs for presence of body fluids from mRNA assay data in mixtures}
\author{R.J.F. Ypma$^1$ \and P.A. Maaskant-van Wijk$^2$ \and R.D. Gill$^3$ \and M. Sjerps$^{3,4}$ \and M. van den Berge$^2$}

\date{
	$^1$Division of digital and biometric traces, Netherlands Forensic Institute \\ \texttt{\{r.ympa, m.sjerps\}@nfi.nl}\\%
	$^2$Division of human biological traces, Netherlands Forensic Institute \\ \texttt{\{p.maaskant, m.van.den.berge\}@nfi.nl}\\[2ex]%
	$^3$Mathematical institute, Leiden University \\ \texttt{gill@math.leidenuniv.nl}\\[2ex]%
	$^4$Korteweg-de Vries Institute for Mathematics, University of Amsterdam \\ \texttt{m.j.sjerps@uva.nl}\\[2ex]%
	18 February, 2020
}

\begin{document}
	\maketitle
	
	\begin{abstract}
		Messenger RNA (mRNA) profiling can identify body
fluids present in a stain, yielding information on what activities could
have taken place at a crime scene. To account for uncertainty in such
identifications, recent work has focused on devising statistical models
to allow for probabilistic statements on the presence of body fluids. A
major hurdle for practical adoption is that evidentiary stains are
likely to contain more than one body fluid and current models are
ill-suited to analyse such mixtures. Here, we construct a likelihood
ratio (LR) system that can handle mixtures, considering the hypotheses
H1: the sample contains at least one of the body fluids of interest (and
possibly other body fluids); H2: the sample contains none of the body
fluids of interest (but possibly other body fluids). Thus, the LR-system
outputs an LR-value for any combination of mRNA profile and set of body
fluids of interest that are given as input.

The calculation is based on an augmented dataset obtained by \emph{in
silico} mixing of real single body fluid mRNA profiles. These digital
mixtures are used to construct a probabilistic classification method (a
`multi-label classifier'). The probabilities produced are subsequently
used to calculate an LR, via calibration. We test a range of different
classification methods from the field of machine learning, ways to
preprocess the data and multi-label strategies for their performance on
\emph{in silico} mixed test data. Furthermore, we study their robustness
to different assumptions on background levels of the body fluids. We
find logistic regression works as well as more flexible classifiers, but
shows higher robustness and better explainability. We test the system's
performance on lab-generated mixture samples, and discuss practical
usage in case work.
\bigskip

\end{abstract}

\noindent\textbf{Note}: this is a prepublication version. The final version appeared in: 
\emph{Forensic Science International: Genetics} \textbf{52}, 2021, 102455,
\url{https://doi.org/10.1016/j.fsigen.2020.102455}.\\
(\url{https://www.sciencedirect.com/science/article/pii/S1872497320302271})

\raggedbottom

\section{Introduction}

In forensic case work, DNA profiling is an important and frequently used
tool, as it has the ability to reveal the identity of a donor of a trace
with high evidentiary value. It is increasingly questioned how
evidentiary traces got deposited (rather than by whom), resulting in
activity level evaluations. Body fluid(s) contributing to an evidentiary
trace can provide such activity level information. Conventional,
presumptive methods for body fluid inference include chemical, enzymatic
and histological assays that tend to be of limited sensitivity and
specificity (both towards body fluids and human origin) and are often
presumptive in nature and only suitable for identification of one body
fluid at a time
\href{https://paperpile.com/c/A0Oh7K/2Pxb+sPWH}{{[}1,2{]}}. Alternative
methods for body fluid inference include messenger RNA-based approaches
\href{https://paperpile.com/c/A0Oh7K/gFjo+7Fhk+ghVP+4YSS+bBbr+z5VP+sQry+D9Sc+NRMs}{{[}3--11{]}}.
These can be targeted at organ tissues or body fluids, and both assays
are applied in forensic casework
\href{https://paperpile.com/c/A0Oh7K/gFjo+8Aqy+zYtd}{{[}3,12,13{]}}.
Inference of organ tissues is less frequent and mainly for objects
involved in violent crimes. The body fluid assay is most frequently
used, mainly in sexual assault cases in which the presence of vaginal
mucosa cells and/or menstrual secretion is disputed. Therefore the most
relevant body fluid to reliably detect are vaginal mucosa and menstrual
secretion. A complicating factor is that many if not all samples will
contain a mixture of human cell types, e.g. due to the omnipresence of
skin cells.

By nature, mRNA markers have neither perfect specificity nor sensitivity
\href{https://paperpile.com/c/A0Oh7K/sPWH}{{[}2{]}}. As a practical
solution, a decision rule or threshold scoring system is used to make a
categorical statement on the presence of body fluids based on the number
of times specific markers were observed
\href{https://paperpile.com/c/A0Oh7K/gFjo+bBbr}{{[}3,7{]}}. These
methods represent ``fall-off-the-cliff'' procedures and to increase
reliability, replicate RNA analysis (generally four) can be applied
\href{https://paperpile.com/c/A0Oh7K/gFjo}{{[}3{]}}. In this approach,
RNA data is evaluated by applying an ``$x = n/2$'' scoring system per body
fluid. Here, ``$x$'' reflects the number of observed and ``$n$'' the number
of theoretically possible signals in all replicates. Body fluids for
which $x \ge n/2$ are reported as ``Indication for the presence of ...''
that body fluid. Body fluids are reported as ``No indication for the
presence of \ldots'' when $x = 0$, and as ``No reliable statement
possible'' when $0 < x < n/2$. This rule has a number
of drawbacks, such as fall-off-the-cliff behaviour (one more marker
detected may completely change the result), an inability to chain
evaluations and an inability to be easily updated when more data becomes
available \href{https://paperpile.com/c/A0Oh7K/aV2v}{{[}14{]}}.
Therefore, there has been increasing interest in statistical methods
\href{https://paperpile.com/c/A0Oh7K/w35J+v1GQ+aQCk+Y94R+I25b}{{[}15--19{]}},
which give a probabilistic statement on the presence of a body fluid in
a sample given measurements. Unfortunately, all methods proposed make
the simplifying assumption that only one body fluid is present per
sample. This makes the methods ill-suited to evaluate mixture data,
which presents a major obstacle to using these methods in practice as
most evidentiary samples may comprise more than one cell type. For
example, experiments in
\href{https://paperpile.com/c/A0Oh7K/aQCk}{{[}17{]}} show that the
method will strongly indicate only one body fluid to be present for
mixture samples, even though the data clearly show multiple body fluids
are present.

Machine learning methods have been gaining traction in a variety of
(forensic) fields, due to their flexibility and ease of use, and have
been proposed for forensic mRNA analysis
\href{https://paperpile.com/c/A0Oh7K/Y94R}{{[}18{]}}. One drawback is
that these models and their predictions are hard to interpret, as the
models often feature many parameters that do not directly correspond to
properties of the modeled system. Furthermore, these classifiers often
are not properly calibrated. An LR system is well calibrated when the
LR-values it produces are not ``too large'' or ``too small'', or more
formally, that the LR of the LRs is the LR. For LR systems based on
machine learning, a post-hoc calibration step is needed. This procedure
may be relatively unknown in forensic biology, possibly because
well-fitting statistical models have long been available for forensic
DNA analysis, but they are routinely used in fields as diverse as
forensic voice \href{https://paperpile.com/c/A0Oh7K/Gv5Y}{{[}20{]}},
glass \href{https://paperpile.com/c/A0Oh7K/X3hh}{{[}21{]}} and cell
phone pattern analysis
\href{https://paperpile.com/c/A0Oh7K/zukR}{{[}22{]}}. Although machine
learning has proved useful in a variety of fields and is now applied to
diverse problems, it is unclear what its utility is in an application
like forensic mRNA analysis that has relatively small datasets.

Another problem facing probabilistic methods in forensic mRNA analysis
is that it is difficult to randomly sample from the (case-dependent)
H\textsubscript{2} population. This random sample is needed to establish
the background levels of body fluids for which H\textsubscript{1} or
H\textsubscript{2} make no specific statements, and thus the markers we
would expect to find if the alleged activity did not take place. For
example, we do not know the background level of saliva in underpants,
nor whether this depends on age, sex or socio-economic background of the
wearer. Fully defining the relevant case-dependent population is
difficult and sampling from it is often infeasible as the population
should be representative with respect to traits that correlate with the
body fluids and markers we would expect to find. Such traits may include
age, socio-cultural background, physiological condition, stress and
sampling site. As statistical models require an explicit assumption on
background levels to be made, previously proposed methods have made the
simplifying assumption of equal background levels. Depending on the case
at hand however, the forensic scientist may feel confident to make
stronger assumptions. For example, we may reasonably assume a penile
sample to include penile skin. Any (probabilistic) method to be used in
practice should not be too sensitive to simplifying assumptions like a
uniform distribution, and should support being updated with relevant
knowledge on background levels by the forensic expert.

In this paper, we construct and evaluate LR-systems based on multi-label
models, i.e. models that explicitly allow for mixtures of body fluids.
We use \emph{in silico} mixing, generating mRNA mixture profiles with a
computer from real single body fluid mRNA profiles. The models include a
range of simple to complex machine learning models, all calibrated to
obtain LRs. We perform a sensitivity analysis to establish the impact of
differing assumptions on background levels, and test performance on mRNA
profiles of mixture samples generated in the lab. We show how background
levels relevant for the case can be included, and discuss practical
usage and interpretation.

\section{Methods \& Materials}

\subsection{RNA profiling data}

The body fluid samples used for this study were collected with informed
consent of the voluntary donors whose cell material was used and the
study was executed according to approved procedures. Body fluid samples
were collected as described by Lindenbergh et al.
\href{https://paperpile.com/c/A0Oh7K/gFjo}{{[}3{]}}. DNA/RNA
co-extraction, DNase treatment, reverse transcription, PCR
amplification, PCR purification and detection for all RNA analyses were
performed according to standardized protocols
\href{https://paperpile.com/c/A0Oh7K/gFjo}{{[}3{]}}. Profile analysis
was performed using Genemapper ID-X version 1.1.1 (Life Technologies)
with a peak detection threshold of 150 relative fluorescence units
(rfu).

We obtained samples for $k=9$ body fluids, and measured expression
of $p=15$ body fluid-specific markers (Table 1) as well as two
control housekeeping markers
\href{https://paperpile.com/c/A0Oh7K/gFjo+7Fhk+8Aqy}{{[}3,4,12{]}}.
These were either processed as single samples, or combined to yield a
mixture sample. Following standard procedures
\href{https://paperpile.com/c/A0Oh7K/7Fhk}{{[}4{]}}, each sample was
analyzed multiple times, yielding two to four replicate measurements per
sample. Samples were discarded if fewer than half the housekeeping
markers amplified. This left 212 single body fluid and 188 mixture
samples of seven body fluid combinations. Note that the single samples
were designed to reflect case conditions, e.g. by including degraded or
low quantity samples, whereas the mixtures were constructed from good
quality samples. This explains the higher detection rates for markers in
the mixture set, i.e. the higher fractions seen in Table 1 (bottom). We
treated the body fluids here as distinct, e.g.~`skin' should be read as
`non-penile skin', and we ignored that menstrual secretion may itself
contain blood or vaginal cells. See discussion for possible improvements
on this simplification.

All data and scripts used are available on github.

\begin{landscape}

\noindent\emph{Table 1 Detection rates of markers, i.e.~the proportion of
replicates (generally four times the number of samples) in which each
marker was detected, per single body fluid (top) and mixture (bottom).
Note: CYP2 stands for CYP2B7P1}
{\tiny
\begin{longtable}[]{@{}llllllllllllllll@{}}
\toprule
\endhead
\textbf{Body fluid(\#samples)} & \textbf{HBB} & \textbf{ALAS2} &
\textbf{CD93} & \textbf{HTN3} & \textbf{STATH} & \textbf{BPIFA1} &
\textbf{MUC4} & \textbf{MYOZ1} & \textbf{CYP2} & \textbf{MMP10} &
\textbf{MMP7} & \textbf{MMP11} & \textbf{SEMG1} & \textbf{KLK3} &
\textbf{PRM1} \\
\textbf{Blood(31)} & 1.000 & 0.960 & 0.579 & 0.000 & 0.000 & 0.000 &
0.000 & 0.000 & 0.000 & 0.000 & 0.000 & 0.032 & 0.000 & 0.000 & 0.000 \\
\textbf{Menstrual secr.(28)} & 1.000 & 0.496 & 0.451 & 0.000 &
0.009 & 0.000 & 0.566 & 0.531 & 0.310 & 0.319 & 0.381 & 0.558 & 0.000 &
0.000 & 0.000 \\
\textbf{Nasal mucosa(31)} & 0.008 & 0.000 & 0.440 & 0.008 & 0.976 &
0.504 & 0.616 & 0.016 & 0.016 & 0.000 & 0.008 & 0.024 & 0.024 & 0.000 &
0.000 \\
\textbf{Saliva(30)} & 0.165 & 0.010 & 0.029 & 0.913 & 0.903 & 0.019 &
0.010 & 0.019 & 0.010 & 0.000 & 0.010 & 0.000 & 0.000 & 0.000 & 0.000 \\
\textbf{Semen fertile(24)} & 0.011 & 0.011 & 0.000 & 0.000 & 0.000 &
0.011 & 0.011 & 0.000 & 0.000 & 0.000 & 0.000 & 0.000 & 0.832 & 0.789 &
0.958 \\
\textbf{Semen sterile(7)} & 0.000 & 0.000 & 0.000 & 0.000 & 0.000 &
0.000 & 0.000 & 0.000 & 0.000 & 0.000 & 0.000 & 0.036 & 0.929 & 0.750 &
0.000 \\
\textbf{Skin(18)} & 0.264 & 0.014 & 0.111 & 0.000 & 0.083 & 0.028 &
0.194 & 0.056 & 0.000 & 0.000 & 0.028 & 0.000 & 0.000 & 0.000 & 0.000 \\
\textbf{Skin penile(12)} & 0.146 & 0.000 & 0.042 & 0.000 & 0.000 &
0.000 & 0.333 & 0.021 & 0.000 & 0.021 & 0.021 & 0.042 & 0.000 & 0.000 &
0.104 \\
\textbf{Vaginal mucosa(31)} & 0.009 & 0.000 & 0.157 & 0.000 & 0.000 &
0.000 & 0.922 & 0.722 & 0.557 & 0.000 & 0.043 & 0.009 & 0.000 & 0.000 &
0.000 \\
\bottomrule
\end{longtable}

\begin{longtable}[]{@{}llllllllllllllll@{}}
\toprule
\endhead
\textbf{Saliva+Semen fertile (32)} & 0.438 & 0.000 & 0.000 & 1.000 &
1.000 & 0.000 & 0.531 & 0.000 & 0.000 & 0.000 & 0.156 & 0.000 & 0.719 &
0.844 & 1.000 \\
\textbf{Semen fertile+ Vaginal mucosa (32)} & 0.000 & 0.000 & 0.031 &
0.000 & 0.000 & 0.000 & 1.000 & 1.000 & 0.875 & 0.000 & 0.094 & 0.000 &
1.000 & 0.938 & 1.000 \\
\textbf{Blood+Nasal mucosa (30)} & 1.000 & 1.000 & 0.967 & 0.000 & 1.000
& 0.467 & 0.733 & 0.000 & 0.100 & 0.000 & 0.000 & 0.000 & 0.167 & 0.000
& 0.000 \\
\textbf{Blood+Vaginal mucosa (14)} & 1.000 & 1.000 & 0.714 & 0.000 &
0.000 & 0.000 & 1.000 & 0.571 & 0.071 & 0.000 & 0.000 & 0.000 & 0.000 &
0.000 & 0.000 \\
\textbf{Blood+Menstrual secretion (16)} & 1.000 & 1.000 & 0.938 & 0.000
& 0.000 & 0.000 & 1.000 & 1.000 & 0.500 & 0.688 & 1.000 & 0.875 & 0.000
& 0.000 & 0.000 \\
\textbf{Saliva+Vaginal mucosa (32)} & 0.000 & 0.000 & 0.531 & 1.000 &
1.000 & 0.000 & 1.000 & 1.000 & 1.000 & 0.000 & 0.125 & 0.000 & 0.000 &
0.000 & 0.031 \\
\textbf{Nasal mucosa+ Saliva (32)} & 0.125 & 0.000 & 0.406 & 0.969 &
1.000 & 0.656 & 0.594 & 0.000 & 0.000 & 0.000 & 0.000 & 0.000 & 0.000 &
0.000 & 0.000 \\
\bottomrule
\end{longtable}

}

\bigskip

\bigskip

\noindent\emph{Table 2. Measurement values for four replicates in three example
cases.}

{\tiny

\begin{longtable}[]{@{}llllllllllllllll@{}}
\toprule
\endhead
\textbf{Finds:} & \textbf{Blood} & \textbf{Saliva} &
\textbf{Saliva/nasal} & \textbf{Nasal} & \textbf{Vaginal} &
\textbf{Menstrual} & \textbf{Semen} & & & & & & & & \\
& \textbf{HBB} & \textbf{ALAS2} & \textbf{CD93} & \textbf{HTN3} &
\textbf{STATH} & \textbf{BPIFA1} & \textbf{MUC4} & \textbf{MYOZ1} &
\textbf{CYP2} & \textbf{MMP10} & \textbf{MMP7} & \textbf{MMP11} &
\textbf{SEMG1} & \textbf{KLK3} & \textbf{PRM1} \\
\textbf{Case 1} & 3/4 & 4/4 & 4/4 & 0/4 & 0/4 & 0/4 & 0/4 & 0/4 & 0/4 &
0/4 & 0/4 & 0/4 & 0/4 & 0/4 & 0/4 \\
\textbf{Case 2} & 4/4 & 4/4 & 4/4 & 0/4 & 0/4 & 0/4 & 4/4 & 4/4 & 4/4 &
4/4 & 4/4 & 4/4 & 0/4 & 0/4 & 0/4 \\
\textbf{Case 3} & 4/4 & 4/4 & 4/4 & 0/4 & 0/4 & 0/4 & 2/4 & 0/4 & 0/4 &
1/4 & 2/4 & 2/4 & 0/4 & 0/4 & 0/4 \\
\bottomrule
\end{longtable}

}

\end{landscape}

\section{Data splitting and augmentation}

We randomly split the single body fluid samples into 40\% `training
data', 40\% `calibration data' and 20\% `test data' before further
processing, stratified by body fluid. The training data are used to fit
a model, i.e., calculate the model parameters. After the training step,
we have a model that for observations on a new sample produces a number
(a `score'). The idea is that samples for which H1 is true will have
higher scores than samples for which H2 is true. Subsequently, the
calibration data are used to calibrate the score, i.e., transform the
score to an LR, which is interpreted as the strength of the evidence.
Thus, we have built an `LR system' that for a new sample calculates the
LR. Finally, the test data are used to test how well this LR system
works.

For each of the training, calibration and test datasets we applied data
augmentation to generate 10, 10 and 5 samples respectively for each of
the $2^k$ possible combinations of body fluids. By data
augmentation we mean the process of creating synthetic data representing
mixtures of body fluids \emph{in silico}, using mRNA marker data from
real samples of single body fluids, see below and Figure 1 for details.
\emph{k}=8 for most analyses, as we exclude penile skin (see methods
section), yielding a total of $2^8 \times (10+10+5)=6400$
augmented samples.

Our data augmentation scheme is as follows. For each of the body fluids
simulated to be in the mixture, we selected a single sample of that body
fluid at random (with replacement) from the dataset of single samples.
We then shuffled the replicates for the selected samples independently,
and constructed augmented replicates by combining the first replicate
for each selected sample, the second, etc. An augmented sample was
constructed from a combination of augmented replicates by taking as peak
height per marker the maximum peak height observed in the combined
replicates for that marker (Figure 1). This reflects our assumption that
mRNA signals for a marker specific for a body fluid will be little
affected by the presence of another non-target body fluid, and that
responses of a marker with non-target body fluids is mostly experimental
noise and will not increase when multiple body fluids are present for
which the marker is not-responsive. Experiments with taking the sum of
the peak heights rather than the maximum (not shown here) give similar
results. Lastly, for each marker, we took the average peak height over
the augmented replicates to be the peak height for that marker for the
augmented sample. We explored the performance of our LR system both on
dichotomized and non-dichotomized data. We obtained the dichotomized
data by setting the value for a marker to 1 if the peak height $\ge 150$
rfu, 0 otherwise.

\noindent\includegraphics[width=\linewidth]{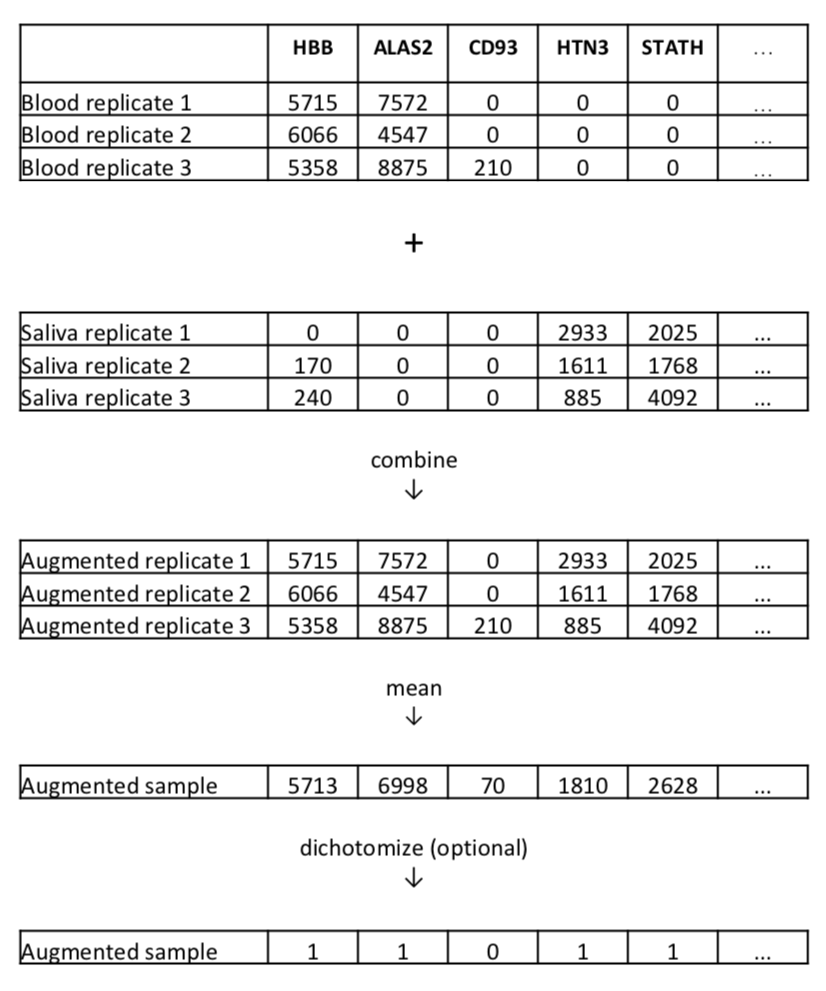}

\noindent \emph{Figure 1. Schematic overview of the augmentation process to
generate a mixture sample (in this case blood and saliva) from single
body fluid mRNA profiles for which three replicates were present.}

\bigskip

Note that the augmented samples are not all independent, which increases
the risk of overfitting of statistical models using the training data.
This is ameliorated by the calibration step, which is performed on the
calibration data independent from the training data. Finally, we test on
augmented test data independent from the training and calibration data,
and on the lab-generated mixture samples that were not used in the model
construction process. The latter is also a good test of our assumptions
in the augmentation process. We repeat this whole process, including the
random splitting, 10 times and evaluate the metrics resulting from these
10 runs.

\section{Methods}

For mRNA data, formulation of the two competing hypotheses differs from
classical `the source of the DNA is person X'/ `the source of the DNA is
a random person (unrelated to X) from the population' for two reasons.
Firstly, rather than having a `reference' individual and a background
population from which random sampling is possible, we have \emph{k}
distinct `individuals' (i.e. body fluids) which we assume to
comprehensively encompass the possible sources, and little knowledge on
background levels. Secondly, the sample of interest most likely results
from a combination of these sources. A subset of the \emph{k} body
fluids will be forensically relevant. The pair of hypotheses we
considered are thus:

\begin{quote}
H${}_1$: the sample contains at least one body fluid of interest (and
possibly other body fluids),

H${}_2$: the sample contains none of the body fluids of interest (but
possibly other body fluids).
\end{quote}

Note that the presented framework could be extended to allow a more
specific defense specification, such as H${}_2$: `the sample
contains body fluid X, none of the body fluids of interest, and possibly
other body fluids'. This could be achieved by only generating samples in
the data augmentation step that are consistent with either H${}_1$, H${}_2$ or
both. See supplementary text for more details.

As background levels for the various body fluids are generally unknown,
we assume a simple uniform distribution. Note that this is not needed
for body fluids for which prosecution and defense agree on
presence/absence, or for which prosecution and defence make no statement
but the forensic scientists feels confident to make a presence/absence
assumption. Blood may be an example of the first, as in some cases it is
clearly visible, penile skin an example of the latter, as it can be
assumed to be present or absent depending on whether we are considering
a penile sample. Thus, we set the background level to 0 for penile skin,
and all other body fluids to be present in 0.5 of all samples,
independently. Practically, we implemented these background levels by
generating mixture samples in the appropriate proportions using the data
augmentation scheme described above. We tested the impact of the
assumption of uniformity using a sensitivity analysis (see below).

To allow for the simultaneous presence of different body fluids, as seen
in mixtures, we constructed a multi-label classifier. Like the more
common multi-class classifier, a multi-label classifier predicts the
presence of each of a number of classes. The difference is that the
former requires the probabilities to sum to one, i.e. makes the hard
assumption that exactly one class is present, whereas in the latter
there is no such restriction, allowing modelling of the simultaneous
presence of zero to all classes. Because of the higher number of
possible outcomes, multi-label classification is often considered a
harder problem than the `standard' multi-class classification. For
multi-label classification an adaption of `standard' classifiers is
usually used \href{https://paperpile.com/c/A0Oh7K/Ijw7}{{[}23{]}}. We
explored two such adaptation strategies: the label power-set method and
one-vs-rest adaptation strategy (see Fig 2). In the former, each of the
2\textsuperscript{k} mixture classes is treated as a new class. Although
accurately calculating the probabilities for each of these classes is
clearly very challenging, the estimates have to be less precise as we
subsequently take the marginal distribution of the set of body fluids of
interest. In the example of figure 2C, this means we would sum the
probabilities of all urns containing blue, arriving at 0.96. In the
latter, the multi-label problem is transformed into a binary
classification problem for the set of body fluids of interest, taking
the marginal distribution during training rather than after. In the
example of figure 2B, blue would represent the set of body fluids of
interests, arriving at 0.9. This strategy has the general drawback that
interaction effects cannot be modelled, which may make this adaptation
strategy less desirable if actual mixture mRNA profiles are available.

\noindent\includegraphics[width=\linewidth]{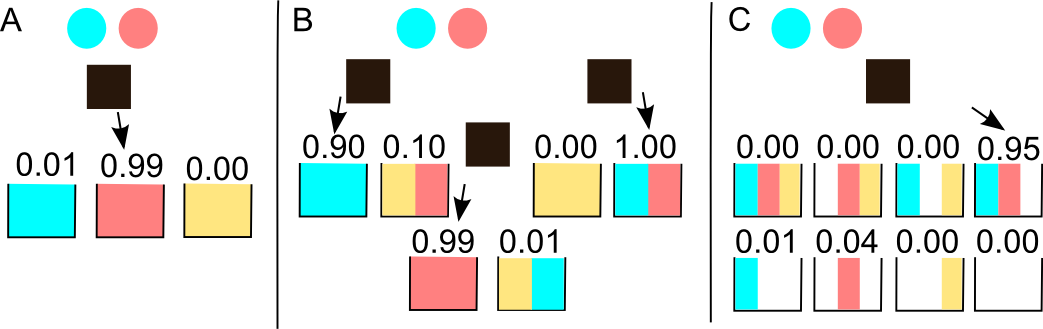}

\noindent \emph{Figure 2. Illustration of different classifiers (black boxes)
having to classify a fictitious sample containing red and blue (discs),
where possible labels are blue, red and yellow (colored urns). (A) A
multi-class classifier cannot correctly classify a sample containing
both red and blue, high certainty on red may artificially lower its
prediction of blue. (B) A multi-label classifier using the one-vs-rest
adaptation strategy consists of three separate binary classifiers, each
predicting the presence of one color, ignoring the other two. (C) A
multi-label classifier using the label power- set adaptation strategy
extends the classes to all possible combinations of original labels, and
uses a multi-class classifier on this expanded space.}

\section{Models}

We explore five classification models (`classifiers') common in the
machine learning literature: (multinomial) logistic regression (MLR)
\href{https://paperpile.com/c/A0Oh7K/w35J}{{[}15{]}}, multi-layer
perceptron (MLP) \href{https://paperpile.com/c/A0Oh7K/6g1L}{{[}24{]}},
support vector machine (SVM)
\href{https://paperpile.com/c/A0Oh7K/ZSpg}{{[}25{]}}, extreme gradient
boosting (XGB)
\href{https://paperpile.com/c/A0Oh7K/gaqz+8zdu}{{[}26,27{]}} and random
forest (RF) \href{https://paperpile.com/c/A0Oh7K/Pj4Z}{{[}28{]}}. These
models are explained below. All of these models provide, after fitting
on training data, a score for new samples. This score is a single
number. For many machine learning models, e.g. SVM, this score has no
probabilistic interpretation but is simply a (transformation of a)
distance measure. For models like MLR the probabilistic interpretation
exists, but can only be taken at face value if we are willing to make
the assumptions imposed by the model, such as independence between
marker values. We thus implement the extra step of calibration, in which
the score distributions on independent data under H\textsubscript{1} and
H\textsubscript{2} are used to map scores to LRs (see section
\emph{Calibration} below). We use the \emph{Python} implementation in
the \emph{xgboost} package
\href{https://paperpile.com/c/A0Oh7K/8zdu}{{[}27{]}} for XGB, and the
\emph{sklearn} package
\href{https://paperpile.com/c/A0Oh7K/ZSpg+USmn}{{[}25,29{]}} for the
others.

\emph{Logistic regression} is a well-known model in both the statistical
and machine learning literature. In the simple case of having to
calculate the posterior probability $P$ on the presence of a single
body fluid, the logistic regression model computes the log posterior
odds as the weighted sum of the measured rfu:
$$\log_{10} \{ P/(1 - P)\}\  = \ {\beta_{0}\  + {\Sigma_{i = 1}^{p}}^{}\beta}_{i}r_{i}\eqno(1)$$
where $p$ is the number of markers, the observations $r$ are
the (possibly dichotomized) rfu values, and the coefficients $\beta$ are
parameters to be estimated. If no markers were detected, the resulting
value would be equal to $\beta_{0}$. The coefficients can be positive
or negative, indicating that detecting the associated marker increases
or decreases our belief that the body fluid of interest is present. Note
that for this model, the posterior probability $P$ that the body
fluid is present is given by the logistic function (where we use base 10
for consistency):

$$P = 1/(1 + 1{0^{}}^{\beta_{0}\  + \Sigma_{i = 1}^{p}\,\beta_{i}r_{i}})$$

These posterior probabilities are based on the assumption that the prior
probabilities are given by the composition in the training data: if the
body fluid is present in 50\% of the training data, then the prior
probability is implicitly taken as 50\%. In this case, the prior odds
are 1 so that the posterior odds equal the LR.

When there are multiple, mutually exclusive body fluids that could be
present, the model can be extended to multinomial logistic regression
(MLR). Unfortunately, the equations for the LRs are no longer as neat,
with the posterior probabilities $P_m$ for the
$k$ body fluids given as

$$P_{1} = 1/(1 + \Sigma_{j = 2}^{k}1{0^{}}^{{\beta_{0,j} + \Sigma_{i = 1}^{p}}\,\beta_{i,j}r_{i}})$$

$$P_{m} = 10^{{\beta_{0,m}\  + {\Sigma_{i = 1}^{p}}^{}\beta}_{i,m}r_{i}}/(1 + \Sigma_{j = 2}^{k}10^{\beta_{0,j} + \Sigma_{i = 1}^{p}\,\beta_{i,j}r_{i}})$$
for $1 < m < k$, 
where $\beta_{i,j}$ is the parameter for the $i$-th marker and
$j$-th body fluid, $0<i\le p$,
$1<j\le k$, resulting in $p(k-1)$
parameters. This is the model suggested in de Zoete et al.
\href{https://paperpile.com/c/A0Oh7K/w35J}{{[}15{]}}. We extend on this
by allowing multiple body fluids to be present simultaneously, using the
one-vs-rest and label power-set strategies introduced above. In the
one-vs-rest strategy, we construct a binary model for the set of body
fluids of interest, i.e. a logistic regression model. This has the
advantage of high interpretability: for prior odds of 1, the log LR is a
linear combination of the estimated parameters (equation 1). We
illustrate this interpretability by showing that these coefficients
behave as expected at the end of the results section (Fig 8). In the
label power-set strategy, we use the same multinomial model defined
above, but consider as classes the 2\textsuperscript{k} possible
combinations of body fluids. To obtain the model output for a set of
body fluids of interest, we sum the probabilities for classes that
contain one or more of the body fluids of interest (H1 is true), and
divide by the sum of the probabilities for classes that contain none of
the body fluids of interest (H2 is true). This ratio is the score that
is produced by the MLR model based on the measured rfu of a new sample.

A multi-layer perceptron (MLP) is an example of an artificial neural
network, a class of statistical models loosely based on the working of
the human brain \href{https://paperpile.com/c/A0Oh7K/6g1L}{{[}24{]}}.
Neural networks have become very popular in machine learning due to good
performance on long-standing difficult problems such as image and text
analysis. A neural network typically consists of several layers of
`nodes', analogous to neurons in the brain. The MLP we use has two such
layers, and the nodes themselves are logistic regression models. The
first layer consists of 100 such models, each taking the features of the
sample (i.e., the measured rfu) as input. The second layer takes the
output of the nodes in the first layer as input. The output of the last
layer is normalised to form the output of the MLP. The number of nodes
in the output layer is equal to the number of classes,
2\textsuperscript{k} for the label power-set adaptation strategy or
\emph{k} for the one-vs-rest adaptation strategy. Note that the second
layer in itself forms a multinomial logistic regression model, working
on the output of the first layer, a `non-normalised' multinomial
regression model. Thus, the MLP is an extension of logistic regression.
For the MLP, the one-vs-rest strategy does not result in independent
models as the first layer is shared for all \emph{k} classes. This means
characteristics of the data do not have to be relearned for each class.
Thus, a model is built that calculates a score based on the measured rfu
of a new sample.

In machine learning, the vector representing the observations is called
the `feature vector'. Here, the feature vector is the
\emph{p}-dimensional vector representing the measured rfu values of the
\emph{p} markers. We can imagine these vectors as points in a
\emph{p}-dimensional space, the `feature space'. In graphical
representations, the composition of the various body fluids in each
sample (the `class' of each sample) can be represented by the colours of
the points. A support vector machine (SVM) defines a hyperplane in the
(possibly transformed) feature space, aiming to separate samples of
different classes and maximize the minimum distance (=margin) from any
sample to the plane
\href{https://paperpile.com/c/A0Oh7K/ZSpg}{{[}25{]}}. The hyperplane can
thus be defined in terms of the feature vectors of the samples that lie
closest to it; the support vectors (i.e. the samples that are closest to
the other class). We use \emph{sklearns} implementation defaults. This
means the one-vs-one adaptation strategy is used to convert this binary
classifier into a multi-class classifier. This strategy is similar to
the one-vs-rest strategy explained above, but rather than building
\emph{k} binary classifiers to distinguish between a class and all other
classes, \emph{k(k-1)/2} binary classifiers are built to distinguish
between every pair of classes. Furthermore, a radial basis function is
used as a kernel, and squared L2 regularization is employed to penalize
non-zero parameters. Lastly, the method proposed by Platt
\href{https://paperpile.com/c/A0Oh7K/BHRM}{{[}30{]}}, itself a form of
logistic regression calibration, is used to convert the distance to the
decision boundary to a score. Thus, a score is obtained from the
measured rfu of a new sample.

Both random forests (RF) and extreme gradient boosting (XGB) are
examples of `ensemble methods', where several simple models (decision
trees) are combined to produce a well-performing model
\href{https://paperpile.com/c/A0Oh7K/8zdu+gaqz+Pj4Z}{{[}26--28{]}}. In a
decision tree, splits are made to the dataset iteratively based on a
single feature, maximizing homogeneity of the split datasets. For XGB,
100 trees with a maximum number of three leaves are sequentially fitted
to the residuals of the previously fit trees. For the RF, 100 trees are
fitted in parallel to a bootstrapped version of the dataset, and their
average predictions are taken as the final prediction.

\section{Calibration}

An important aspect to consider for any LR system is whether it is
well-calibrated. Intuitively, a probabilistic system is well-calibrated
when events for which it gives a probability of 0.9 happen 90\% of the
time. For example, if a weather forecasting system's predictions of 0.9
probability of rain are followed by rain only 50\% of the time, we would
call the system, and the probabilities it outputs, uncalibrated. We
could `calibrate' the system, by adjusting the uncalibrated probability
of 0.9 to the calibrated probability of 0.5. Equivalently, in a forensic
context, for well-calibrated systems the property `the LR is the LR of
the LR' holds
\href{https://paperpile.com/c/A0Oh7K/LdyY+GnLs}{{[}31,32{]}}.
Intuitively, this means that the size of the LR is coupled with the
frequency of occurrence under each hypothesis. For example, if we
observe a set of hundred samples taken under H2 conditions, we may
expect to observe a `misleading' LR of ten about ten times, and a
misleading LR of hundred just once. If we observe a misleading LR of one
million three times, we have a clear indication that our LR-system is
calculating LRs that are off, i.e. are not consistent with expected
frequencies. We then say that the LR-system is ill-calibrated.
Ill-calibrated LR-systems are of use if they have discriminative power,
i.e. if LRs calculated for H1 tend to differ from LRs calculated for H2.
The output of any statistical model could be ill-calibrated, for example
if the model assumptions deviate too much from reality.

The machine learning models introduced above have been found in the
machine learning literature to provide good performance to a wide
variety of problems. However, the machine learning literature usually
does not focus on calibration of the output, as performance in many
applications only pertains to the relative sizes of the probabilities,
i.e. discriminative power. We should thus not assume these models to be
well-calibrated. To emphasize that the uncalibrated LRs output by the
models should not be interpreted as calibrated LRs, we refer to them as
scores \href{https://paperpile.com/c/A0Oh7K/JLBY}{{[}33{]}}. In order to
interpret them as LRs, we need to transform them by `calibration'. We
implemented a calibration step in which a logistic regression model is
fitted on the log scores
\href{https://paperpile.com/c/A0Oh7K/7cqN+Gv5Y+aEuf}{{[}20,34,35{]}},
see figure 3 for an illustration. The calibration step is performed for
each set of body fluids of interest separately. Thus, each `score' is
transformed into a number that we now call LR. Fig 3(C) shows the
transformation function from score to LR for an example.

Note that, for uniformity, we also apply this logistic regression step
to the logistic regression model. Luckily, for the one-vs-rest
adaptation strategy, the additional logistic regression calibration step
preserves interpretability, as it only scales and translates the
coefficients:
$$\log_{10}LR~= ~a_0 + a_1({b_{0}\  + \Sigma_{i = 1}^{p} b_{i}r_{i})~= ~{(a_0 + a_1\ {b}_{0})+ \Sigma_{i = 1}^{p}a_1b_{i}r_{i}~ =~ \beta_{0} + \Sigma_{i = 1}^{p}\beta}_{i}r_{i}} \eqno(2)$$

\noindent\includegraphics[width=\linewidth]{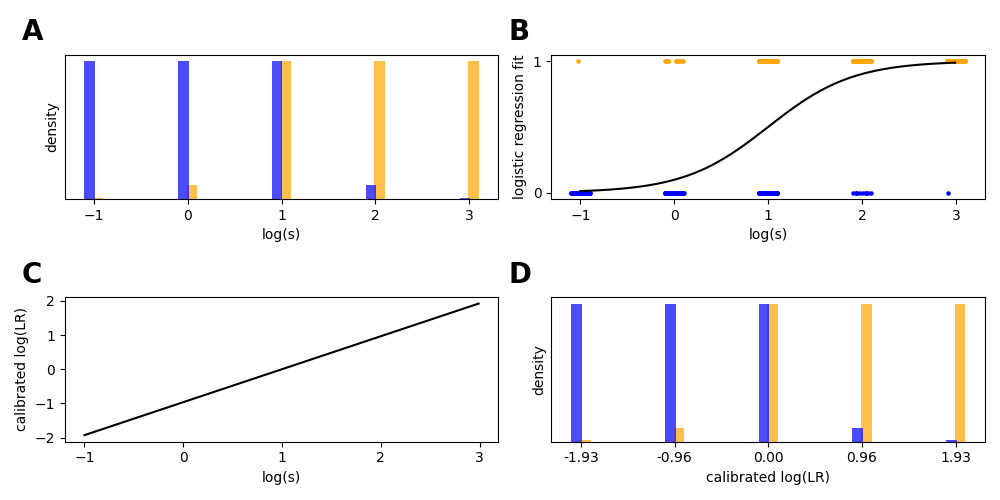}

\noindent\emph{Fig 3. Illustration of calibration using logistic regression}. 

\emph{(A) Distribution of $\log_{10}$ scores $s$ for (orange)
\text{\rm H}${}_1$ and (blue) \text{\rm H}${}_2$ samples, for a
hypothetical model. If the model were well-calibrated, the scores would
be interpretable as LRs. The model is discriminative as the scores
discriminate between the hypotheses, but ill-calibrated as `the LR is
the LR of the LR' does not hold for $s$. For example, there are 10 times
more \text{\rm H}${}_2$ samples that result in $\log_{10}(s)=0$ than
there are \text{\rm H}${}_1$ samples that do so, meaning
$P(s=1| \text{\rm H}_1)/P(s=1|\text{\rm H}_2)=\text{\rm LR}(s=1)=0.1$.
Likewise \text{\rm LR}($s$=0.1)=0.01 and \text{\rm LR}($s$=10)=1. The model would be
well-calibrated if all scores were 10 times smaller.}

\emph{(B) The fit of a
logistic regression model that outputs
P({\rm H}\textsubscript{1}\textbar $\log$(s))=P\textsubscript{log\_reg} from the
distribution in (A). Dots represent the actual data points (some random
noise in the x-axis was added just to be distinguishable).}

\emph{(C) The
mapping from $\log(s)$ to
$\log$(P\textsubscript{$\log$\_reg}/(1-P\textsubscript{$\log$\_reg})) illustrates
that logistic regression gives a linear mapping in the log odds space.
We refer to P\textsubscript{log\_reg}/(1-P\textsubscript{$\log$\_reg}) as
the calibrated {\rm LR}.}

\emph{(D) Distribution for {\rm H}\textsubscript{1}} \emph{and
{\rm H}\textsubscript{2} samples obtained after logistic regression
calibration. Note that the calibration still is not perfect, as values
are shrunk slightly towards $\log$({\rm LR})=1.}

\section{Measuring performance}

When evaluating an LR system and comparing it to others, several aspects
are of interest \href{https://paperpile.com/c/A0Oh7K/wFZV}{{[}36{]}}.
The most obvious is discrimination, the ability of a system to tell
apart the competing hypotheses, as measured by metrics such as accuracy
or area under the curve of the receiver operator characteristic. These
measures however ignore calibration, i.e. whether not just the ordering
but also the magnitude of the LR values is what it should be. We
therefore use the log likelihood ratio cost
(\emph{C\textsubscript{llr}}), the empirical cross-entropy for equal
priors, a metric suggested by information theory that measures both
discrimination and calibration
\href{https://paperpile.com/c/A0Oh7K/3T03}{{[}37{]}}. Its value will be
0 for a perfect system, and 1 for a system that does not improve
decision making (e.g. a system that always outputs LR=1). This means any
system with a value of \emph{C\textsubscript{llr}}\textless1 will
improve decision making, with smaller values being better.

\section{Sensitivity analysis}

As noted before \href{https://paperpile.com/c/A0Oh7K/w35J}{{[}15{]}},
since we are not sampling at random from the population but assuming
equal background levels for the body fluids, a deviation from this
assumption may affect the performance of our LR systems. Although we
cannot establish the background levels, we can explore the magnitude of
this effect by constructing the same models using differing assumptions
on these background levels, and assessing the differences in LR values
calculated. We generated augmented datasets of the same size in which a
particular body fluid is present in 0.9 rather than 0.5 of generated
samples. We train a system on this dataset using the same procedure, and
plot the LRs it generates against the LRs generated by the original
system. A large deviation would indicate a high sensitivity which would
make the method less useful in practice.

In total, we performed 10 runs of different dataset splits, each run
consisting of 2 dichotomization of data x 2 multi-label strategies x 4
models = 16 settings. We show \emph{C\textsubscript{llr}}s for the set
of body fluids of interest being vaginal mucosa and/or menstrual
secretion. We performed the 10x 16 x 2 set of computations 3 more times,
increasing the background levels of blood, nasal mucosa and skin, and
compared the LRs generated by the original model with the LRs generated
by the models trained on these non-uniform data.

\section{Practical use and interpretation}

To illustrate the use and interpretation of the best performing model,
we present three more results. First, we illustrate the best performing
model on illustrative cases, discuss the characteristics of the model,
and how these change when the background level of penile skin is set to
1. Second, we contrast the multi-label model to previously proposed
multi-class models. Third, we assessed performance of the model on 86
traces from RNA casework conducted at our laboratory between 2018 and
2020 for which the presence of vaginal mucosa and/or menstrual secretion
was questioned. These have been analyzed in the context of the case
using the n/2 interpretation method
\href{https://paperpile.com/c/A0Oh7K/gFjo}{{[}3{]}} summarized in the
introduction. We compare conclusions given by the n/2 method and the
model described in this paper.

\section{Results}

Results shown have vaginal mucosa and/or menstrual secretion as body
fluids of interest, and logs are to base 10, unless specifically stated
otherwise.

Performance for all models and multi-label strategies on
dichotomized/non-dichotomized data are shown in figure 4. We see that
under all scenarios tested the methods perform well, with
\emph{C\textsubscript{llr}}s ranging from 0.1 to 0.7. This means that,
under the assumptions of the augmentation scheme, using any of these
methods will improve decision making. Generally, the models perform very
similarly, with distributions of \emph{C\textsubscript{llr}}s over the
10 runs mostly overlapping. For illustration, figure 5 shows the
underlying LR distributions and ROC curves (for the MLR, one-vs-rest,
dichotomized method), for a number of (combinations of) body fluids of
interest.

\noindent\includegraphics[width=\linewidth]{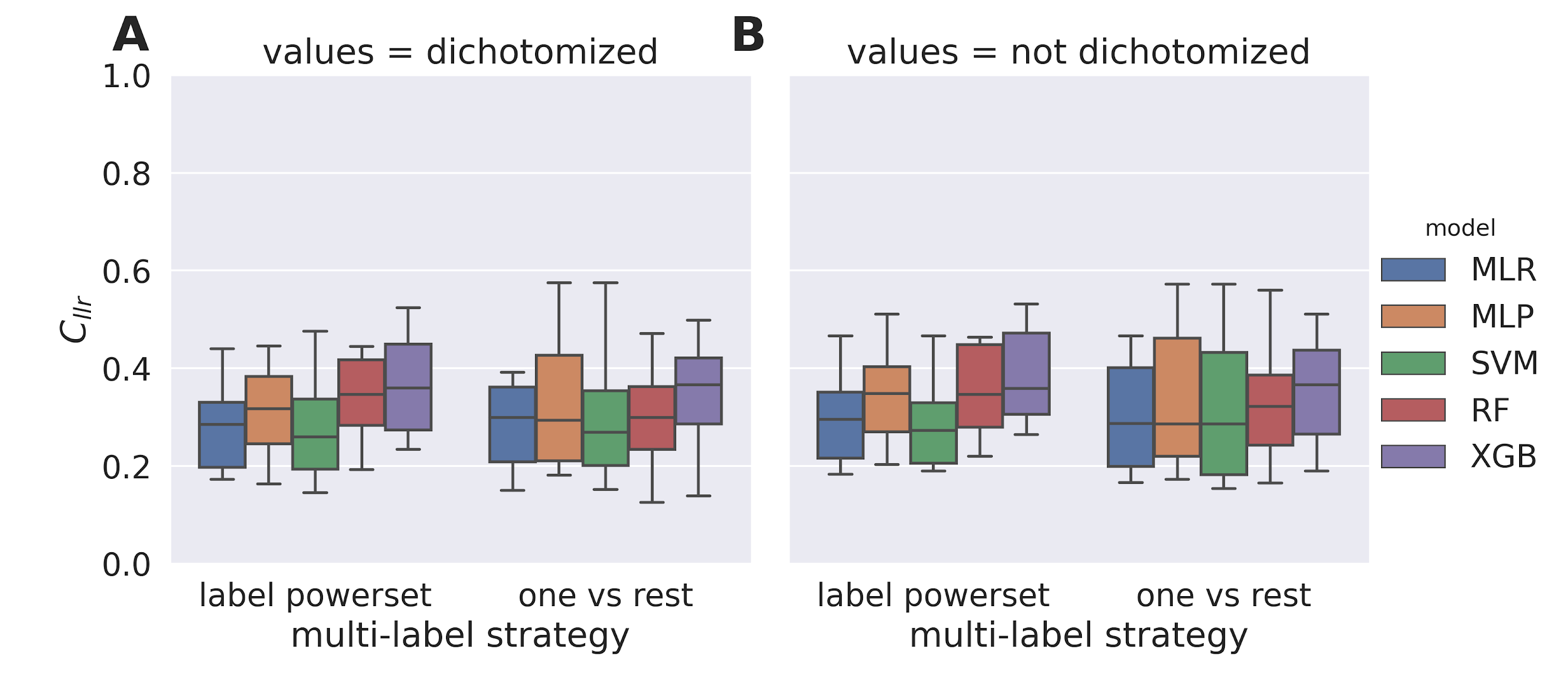}

\noindent \emph{Fig 4. Distributions of C\textsubscript{llr} on the augmented test
data for the different tested scenarios: Data (A) dichotomized or (B)
not; using the `one-vs-rest' strategy (5 boxplots on right) or `label
power-set' method (5 boxplots on left). Box colour marks the five
different classifiers.}

\newpage

\noindent\includegraphics[width=0.5 \linewidth]{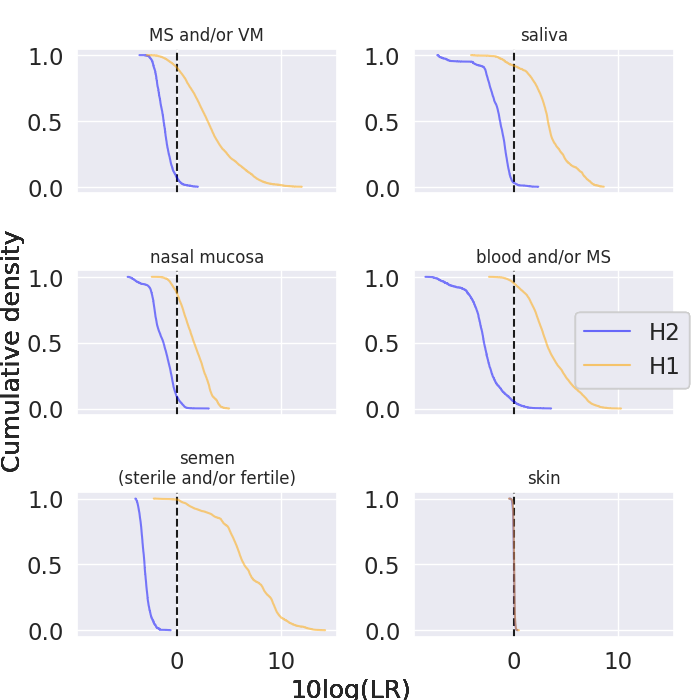}\includegraphics[width=0.5 \linewidth]{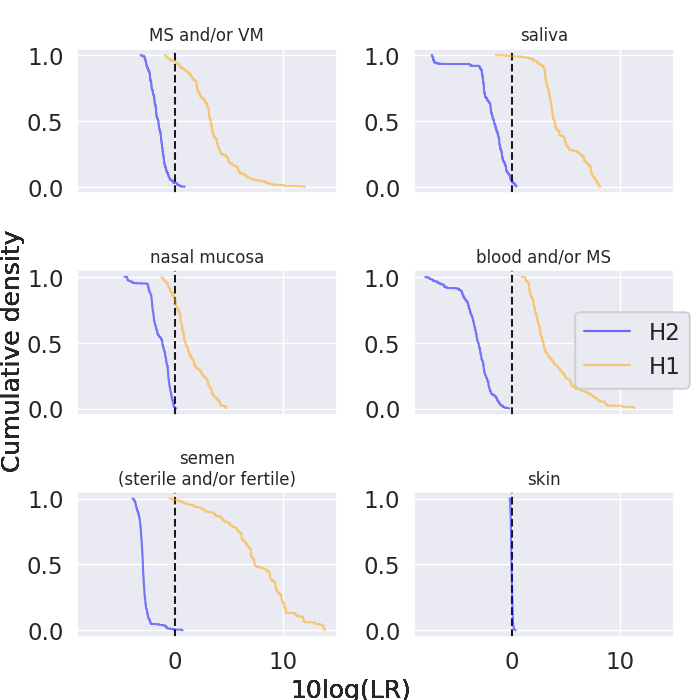}

\noindent\includegraphics[width=0.5 \linewidth2]{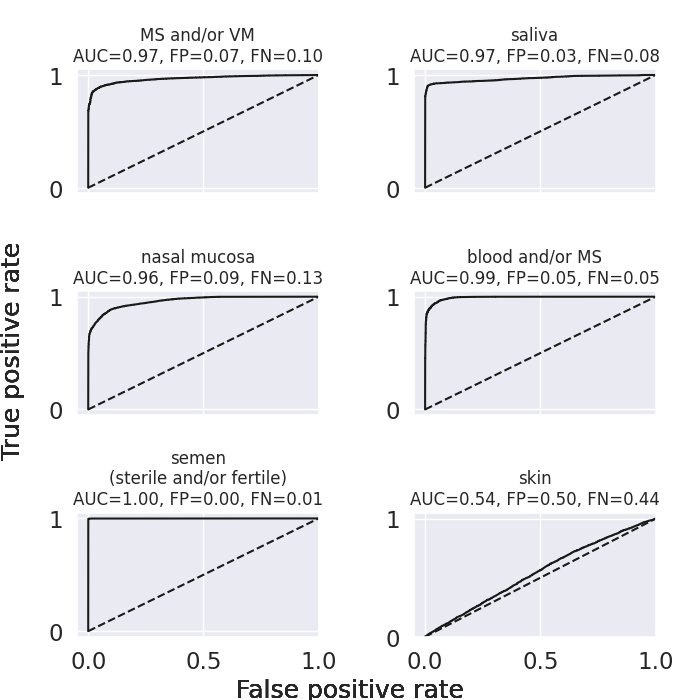}\includegraphics[width=0.5 \linewidth]{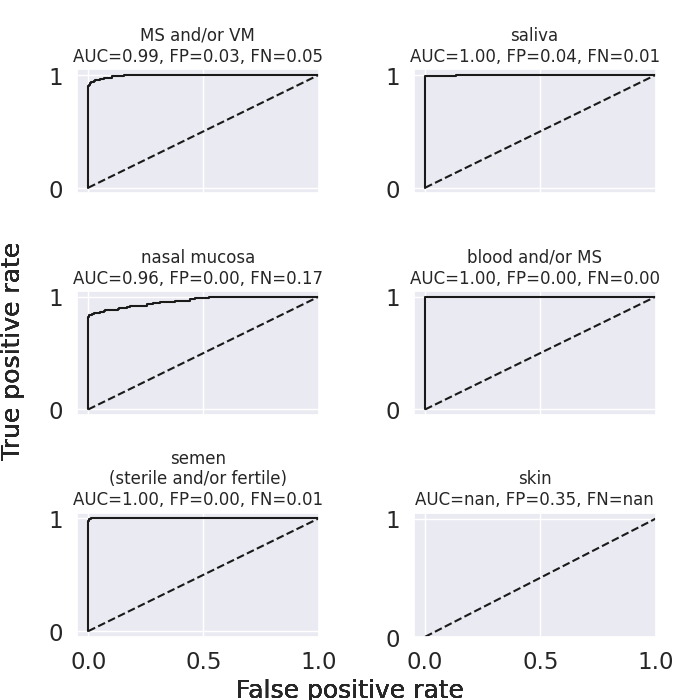}

\noindent \emph{Fig 5. Performance of the MLR, one-vs-rest, dichotomized method,
for six different sets of body fluids of interest, for (A, C) augmented
test data and (B, D) lab-generated test data. A and B show inverse
cumulative density functions of \textsuperscript{10}log LRs (`Tippett
plots') for (orange) H1 and (blue) H2. C and D show receiver operating
curves, reporting on the area under the curve (AUC) and false positive
(FP) and false negative (FN) rates. Note that no skin markers are
present in the data, correctly leading to LR\textasciitilde1 (A; bottom
right) and a curve close to the diagonal (C; bottom right) when skin is
assessed. As the lab-generated mixture samples contain no examples with
skin, B and D bottom right have no H1 (orange) distribution, AUC or FN
rate.}

\bigskip

Figure 6 shows \emph{C\textsubscript{llr}}s when the LR systems are
tested (but not trained/calibrated) on independent lab-generated mixture
data. We note that for the dichotomized data (left panel) all models and
both multi-label strategies perform better than on the augmented test
data (i.e. show lower \emph{C\textsubscript{llr}}). This performance
increase may be explained by the mixture samples being less degraded by
design (see Data section). The similar performance of the two
multi-label strategies on the test data (Fig 4) motivates us to focus
henceforth on the one-vs-rest strategy, as it is the simpler method. In
particular for the MLR model it leads to higher interpretability (see
Interpretability section below).

In contrast, without dichotomization, most models perform worse, even
showing \emph{C\textsubscript{llr }}values above the critical threshold
of 1 (Fig 6, right panel). This poor performance may well be due to
faulty assumptions in the way we generate peak heights in the data
augmentation scheme, or because the peak height distribution is
different for the mixture data that were not designed to reflect case
conditions. Dichotomizing data is less sensitive to actual peak heights,
which would explain the lack of a performance drop. Only the decision
tree-based models (RF and XGB) show reasonable performance, albeit worse
than on the dichotomized data. This is probably because such models use
thresholds to handle non-categorical data, making them inherently robust
to changes in peak heights. In the following we therefore focus on the
dichotomization strategy, for its higher robustness.

\bigskip

\noindent\includegraphics[width=\linewidth]{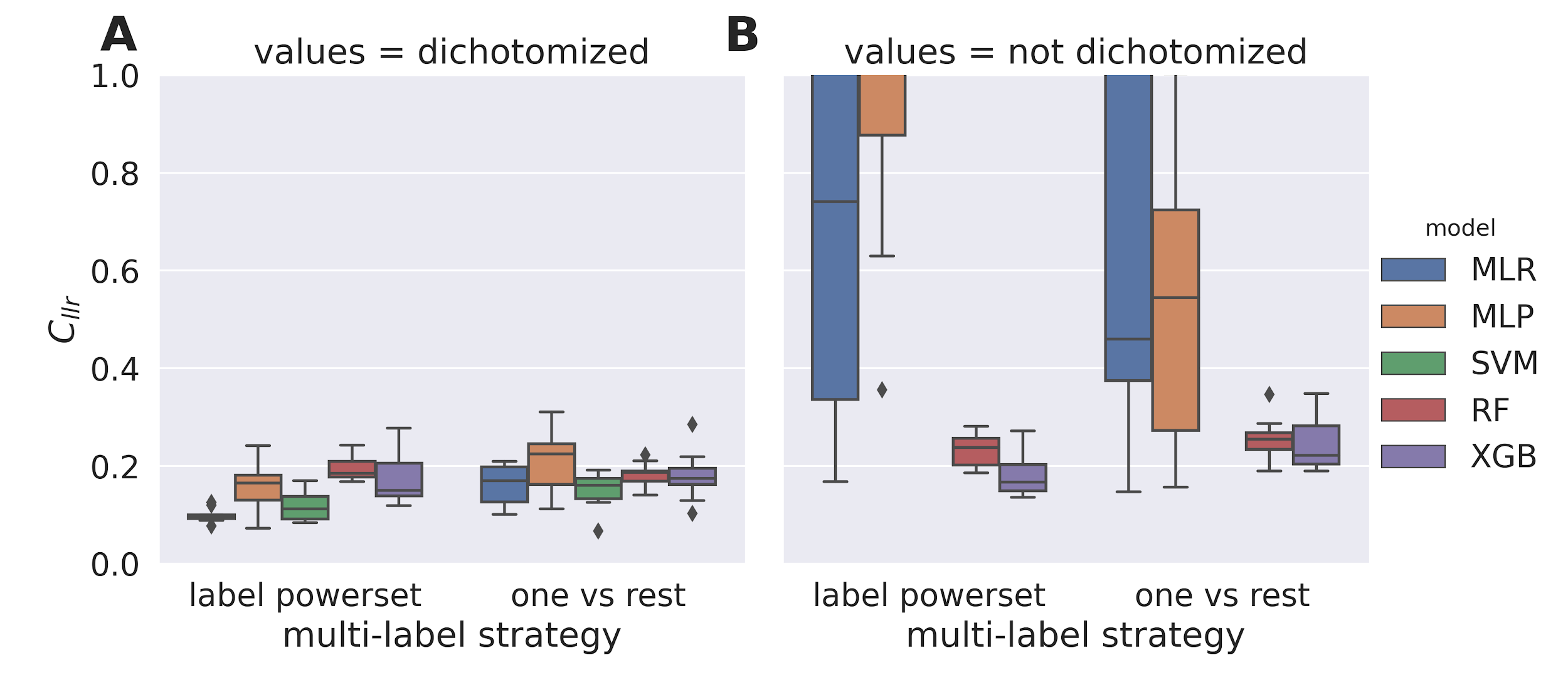}
\noindent\emph{Fig 6. Distributions of C\textsubscript{llr} on the laboratory mixture data
for the different tested scenarios for: Data (A) dichotomized or (B)
not; using the `one-vs-rest' strategy (5 boxplots on right) or `label
power-set' method (5 boxplots on left). Box colour marks the five
different classifiers. Note that in (B), all C\textsubscript{llr}s for
the SVM were \textgreater{} 1.}

\bigskip

\noindent\centerline{\includegraphics[width=0.80\linewidth]{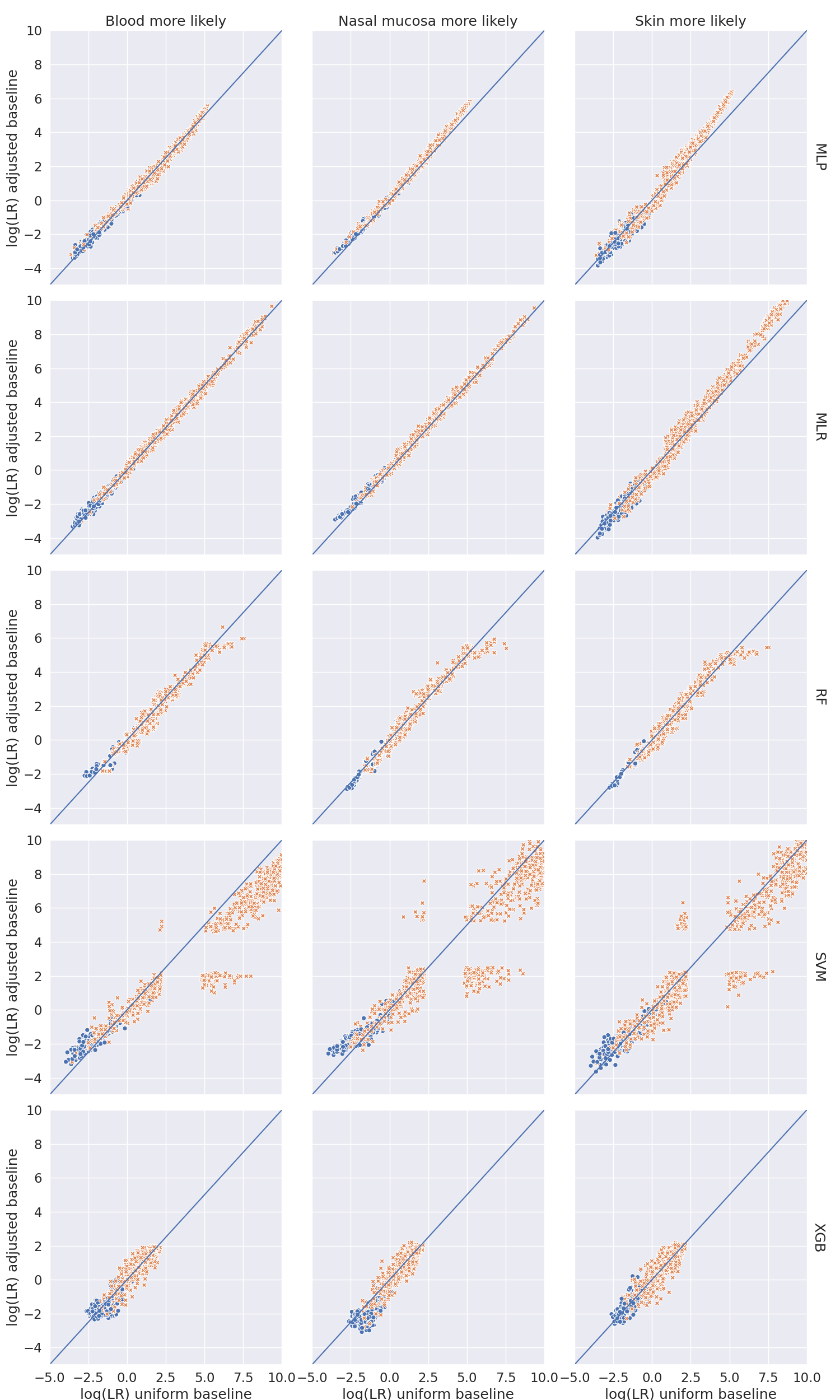}}

\noindent\emph{Fig 7. $\log_{10}\, \text{\rm LR}$s obtained for samples containing
(orange) vaginal mucosa and/or menstrual secretion or (blue) neither of
these. y-axis: training with the assumption of a background level of
90\% probability for the presence of blood, nasal mucosa or skin.
x-axis: training with the assumption of uniform background levels. Rows
show scatterplots for the multi-layer perceptron (MLP), multinomial
logistic regression (MLR), random forest (RF), support vector machine
(SVM) and extreme gradient boosting (XGB) models, using dichotomization
and the one-vs-rest multi-label strategy.}

\newpage

Figure 7 shows the LRs obtained on an augmented test set constructed
with equal background levels, for models trained on augmented train data
with equal background levels (uniform) or a background level of 90\%
probability for the presence of blood, nasal mucosa or skin respectively
(with dichotomization, one-vs-rest strategy). We see that the difference
between LRs obtained ranges from low (MLR and MLP) to very high (SVM).
Small differences in LRs are indicative of the LR system being
relatively insensitive to a violation of the assumption of equal
background levels. We will focus on MLR as its performance is comparable
to other models (Fig 4), its robustness is similar or better (Fig 7) and
its interpretability is much higher (see below).

\section{Practical use and interpretability of the MLR one-vs-rest LR
system}

In this section, we illustrate usage of our preferred model, logistic
regression (MLR) on dichotomized data and a one-vs-rest adaptation
strategy. As logistic regression is linear regression in the log odds
space, the log LR can be calculated as a weighted sum of the measured
markers (see equation 2). To illustrate this, consider three cases in
which the presence of vaginal mucosa and/or menstrual secretion is
disputed, and for which measurements for four replicates are given in
table 2. We can compute the relevant log LRs from equation 2 by plugging
in the coefficients given in figure 8.

For case 1 only the blood markers are observed: 3/4 for HBB and 4/4 for
both ALAS2 and CD93 (Table 2). Although we expect these to be present
when menstrual secretion is present, the absence of menstrual secretion
markers indicates only blood was present in the sample. This is
consistent with the negative log LR given by the model (where the $\beta$'s
are the coefficients in Figure 8 and the $r$'s are the marker
results in Table 2):
$$\log_{10}\text{LR}~ = ~{\beta_{0}\  + \Sigma_{i = 1}^{p}\beta}_{i}r_{i} \hskip5 cm$$
$$\textstyle\hskip2 cm=~-1.34 + 0.79 * \frac34 + (-0.57) * \frac44 + (-0.10) * \frac44 ~=~ -1.4$$
For case 2 all markers that we would expect for vaginal mucosa and/or menstrual secretion are present, the model gives back a higher LR as
expected:
$$\textstyle\log_{10}\text{LR}~ = \hskip5 cm$$
$$\textstyle\hskip2 cm-1.34 + 0.79 * \frac44 + (-0.57) * \frac44 + (-0.10) * \frac44 +
1.45*\frac44 + 1.33*\frac44 + 2.75*\frac44 + 0.56*\frac44 + 1.35*\frac44 + 2.32*\frac44$$ $$\hskip5 cm ~=~ 8.5$$
For case 3 again blood is clearly present, but there is a clear
indication of vaginal mucosa and/or menstrual secretion, with some
expression of the relevant markers. Note that the n/2 method
\href{https://paperpile.com/c/A0Oh7K/gFjo}{{[}3{]}} described in the
introduction would have concluded ``No reliable statement possible''. In
contrast, the MLR model output does reflect this indication:

$$\textstyle\log_{10}\text{\ LR}~=~ -1.34 + 0.79 * \frac44 + -0.57 * 4/4 + -0.10 * \frac44 +
1.45*\frac24 + 0.56*1/4 + 1.35*\frac24 + 2.32*\frac24 = 1.5$$

Knowledge of the case may change this assessment. For example, we know
penile skin sometimes gives weak expression on MUC4, MMP10, MMP7 and
MMP11 (table 1). Thus, if the sample under consideration were a penile
sample, penile skin would form an alternative explanation for the
presence of these markers. Thus, we would expect the log LR (for vaginal
mucosa and/or menstrual secretion) to be lower. We can compute the log
LR taking the penile sampling in account, i.e. by setting the background
level for penile skin to 1. This results in adjusted coefficients (fig
8, bottom), and in a log LR of:

$$\textstyle\log_{10}\text{\ LR}~=~ -1.65 + 0.51 * \frac44 + -0.43 * \frac44 + 0 * \frac44 +
0.81*\frac24 + 0.82*\frac14 + 1.1*\frac24 + 2.4*\frac24 ~=~ 0.8$$
which is lower, as expected.

\bigskip

\noindent\includegraphics[width = \linewidth]{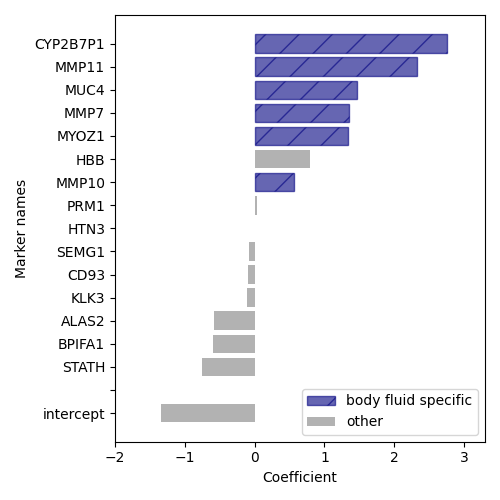}

\noindent\includegraphics[width=\linewidth]{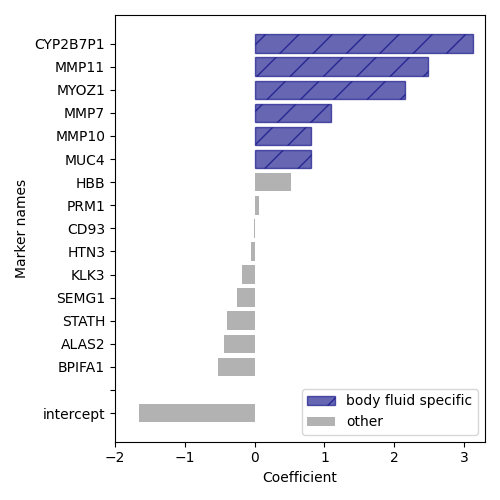}

\noindent \emph{Fig 8. Coefficients and intercept for the various markers for the
logistic regression (MLR) model (dichotomized data, one-vs-rest
strategy). Coefficients shown are relevant (top) when sampling not
penile skin or (bottom) when sampling penile skin, i.e. background level
for penile skin is set to (top) 0\% probability or (bottom) 100\%
probability. Coefficients for markers that have been selected to
indicate vaginal mucosa and menstrual secretion are shown in purple.}

\bigskip

The values found for the coefficients for the model (fig 8) can to some
extent be explained. As expected, coefficients are large for markers
that were originally selected for their sensitivity to menstrual
secretion and vaginal mucosa (fig 8, purple). The coefficient is also
large for HBB; this is a marker selected for sensitivity to blood that
is expressed for all samples with menstrual secretion measured (see
table 1). There are three markers with strong negative coefficients,
STATH, BPIFA1 and ALAS2. These markers are expressed in nasal mucosa and
blood, which are the body fluids hardest to distinguish from vaginal
mucosa and menstrual secretion
\href{https://paperpile.com/c/A0Oh7K/8Aqy}{{[}12{]}}. Thus, detection of
these markers increases the likelihood of the sample containing nasal
mucosa and/or blood. Although the presence of these body fluids in
itself holds no information on the presence of vaginal mucosa and/or
menstrual secretion, it does mean that the presence of vaginal mucosa or
menstrual secretion are no longer needed to explain jointly expressed
markers (e.g. MUC4). Thus seeing BPIFA1 and MUC4 should yield a lower LR
when assessing vaginal mucosa and/or menstrual secretion as body fluid
of interest than just seeing MUC4, which is modelled by MLR as a
negative coefficient. Note that this is actually a more advanced
interaction effect between markers that cannot be modelled by the
logistic regression model, which assumes independence. Lastly, MUC4 and
HBB are sometimes detected in penile skin. Thus, in a sample that we
know to contain penile skin, the presence of these markers should be
less indicative of vaginal mucosa and/or menstrual secretion. This is
exactly what we see reflected in the smaller coefficients for these two
markers for the model with a penile skin background level of 100\% (fig
8 bottom).

\bigskip

\noindent\includegraphics[width=\linewidth]{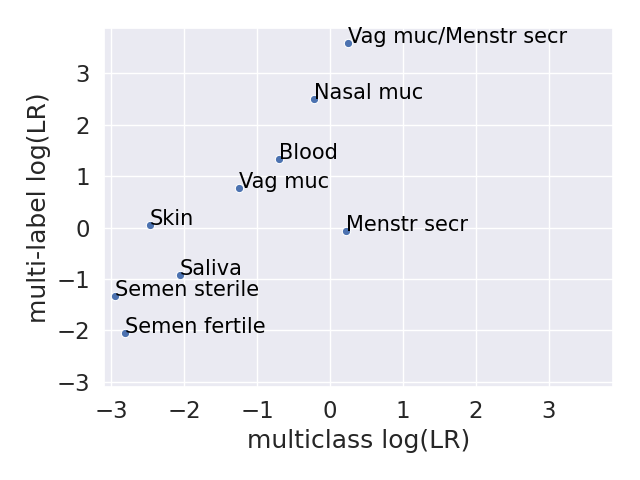}

\noindent\emph{Fig 9. Illustration of model predictions using (x-axis) a
previously proposed multi-class
\href{https://paperpile.com/c/A0Oh7K/w35J}{{[}15{]}} or (y-axis) the
present multi-label logistic regression model, on a sample containing
blood, nasal mucosa and vaginal mucosa.}

\bigskip

Figure 9 illustrates the advantage of using a multi-label model on
mixtures. LRs are plotted for the different body fluids and the
combination vaginal mucosa and/or menstrual secretion for a fictitious
sample containing blood, nasal mucosa and vaginal mucosa. We constructed
the sample to have `perfect' expression, i.e. all relevant markers (HBB,
ALAS2, CD93, STATH, BPIFA1, MUC4, MYOZ1, CYP2B7P1) were expressed for
all replicates. This measurement gives strong support for the presence
of all three body fluids. As expected, figure 9 shows the multi-label
model assigns LR\textgreater1 for each of the three body fluids.
However, multi-class logistic regression
\href{https://paperpile.com/c/A0Oh7K/w35J}{{[}15{]}} assigns
LR\textasciitilde1 to menstrual secretion, and LR\textless1 to all other
body fluids, because by construction it cannot handle mixture data. Note
that the presence of blood and vaginal mucosa markers could also have
been caused by the presence of menstrual secretion. This makes it hard
to say whether vaginal mucosa or menstrual secretion are present, which
leads to the multi-label model assigning LRs close to 1 to both.
However, it is clear from the markers that at least one of the two body
fluids must be present, as is reflected in the high LR
(\textgreater1000) the model assigns to the combined class of vaginal
mucosa and/or nasal mucosa.

Figure 10 gives an overview of conclusions of the n/2 method vs those
from the proposed method, for actual casework. 34 out of the 86 traces
resulted in a ``indication for the presence of vaginal mucosa''
reporting using the n/2 interpretation method. Using the developed
model, all of these traces resulted in LRs ranging from ``moderate
support'' to ``moderately strong support'' according to the ENFSI verbal
scale (model capped to an LR of 1000 \emph{i.e.} ``moderately strong
support''). 23 traces fitted in the category with ``no indication for
the presence of vaginal mucosa'' based on the n/2 method, all except one
of these resulted in an LR \textless0.5 when the model was used (one
sample resulted in an LR of 1.01). The remaining 29 traces fitted the
``No reliable statement possible'' category. For 16 of these traces, the
model resulted in an LR \textless0.5. Two traces resulted in an LR
between 0.5 and 2 (``do not support one hypothesis over the other'') and
eight traces in an LR between 2-10 providing ``weak support''. Three
samples resulted in an LR between 10-100 (``moderate support''). Thus
for these latter 11 traces, the model provided information supporting
the presence of vaginal mucosa and/or menstrual secretion, whereas when
using the n/2 method, this information is lost and these traces were
reported as ``No reliable statement possible''.

\bigskip

\noindent\includegraphics[width=\linewidth]{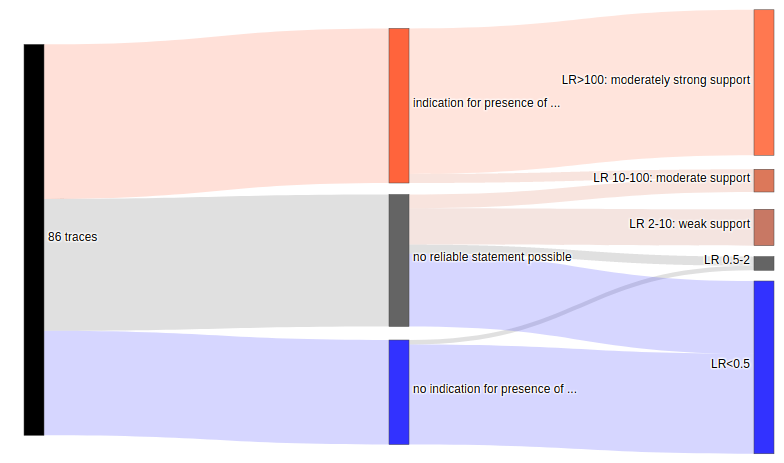}

\emph{Fig 10. Overview of conclusion reached for 86 casework traces for
(middle column) the n/2 method currently in use and (right column) the
proposed logistic regression model.}

\section{Discussion/future work}

We have presented a novel way to probabilistically assess the presence
or absence of forensically relevant body fluids, allowing for the
presence of multiple body fluids in a sample. We constructed and tested
the performance of LR systems based on machine learning models followed
by a calibration step. We found that performance, as measured by
\emph{C\textsubscript{llr}}, was similar for different models and
multi-label strategies. However, the method seemed more robust when
using dichotomized data, showing better performance on a lab-generated
mixture dataset, and when using logistic regression, with less
sensitivity to deviations from the assumption of equal background levels
for the body fluids. Using logistic regression and the one-vs-rest
multi-label strategy yields a highly interpretable model, whose
predicted \textsuperscript{10}log LRs are a weighted summation of its
coefficients. Application to historical casework showed results that
were consistent with those given by the currently used `n/2' evaluation
method, but more informative for 11/86 cases. We therefore recommend
implementing this system in forensic casework, with extra validation
studies needed for different lab environments and when body fluids of
interest differ from vaginal mucosa and menstrual secretion.

The evaluation of presence/absence of body fluids from mRNA data is
difficult as relatively few data are available given the problem
complexity. For example, relevant markers like MYOZ1, CYP2B7P1 and MMP7
are seen exactly once or twice for nasal mucosa samples. If all or none
of the instances of seeing a marker end up in the training set, any
flexible statistical model is likely to over-infer from this that the
marker is highly specific. An order of magnitude more data may well be
needed to benefit from the flexibility such models offer. This may
explain why the previously suggested
\href{https://paperpile.com/c/A0Oh7K/w35J}{{[}15{]}} logistic regression
model, often considered a baseline in machine learning literature,
performs best. Likewise, more sophisticated multi-label techniques that
allow for interactions, such as the here-studied label power set or
not-studied classifier chains
\href{https://paperpile.com/c/A0Oh7K/P2lO}{{[}38{]}}, may only prove
their worth when (more) experimental mixture data are available.

To compute LRs in a multi-class setting information on background levels
of the different classes is needed
\href{https://paperpile.com/c/A0Oh7K/w35J}{{[}15{]}}. These background
levels are not known. In the present study, this information is needed
to set the proportion of body fluids in the \emph{in silico} mixtures
used to construct the LR systems. There are three reasons we think our
assumed background level of 50\% probability for the presence of every
body fluid, which is unrealistic, is nonetheless acceptable. First, it
is a uniform distribution, not giving preference to any of the body
fluids. Second, the assumption leads to conservative LR systems. As 50\%
is probably an overestimate, we obtain more body fluids per sample than
would be expected in practice, which is a harder problem for the models
to solve. The calibration ensures this leads to LRs closer to 1. Third,
the approach allows, for individual cases, to test the impact of
different assumptions by explicitly setting the background levels,
including the extremes of 0\% and 100\%.

Although we have focused here on generic, flexible models, better
fitting models grounded in biology may be obtained with mathematical
modelling of mRNA expression in cells. This would entail describing what
mRNA profiles look like as a function of the body fluids present,
similar to current practice in forensic DNA analysis. Effects of
physiological status, environmental factors and impacts after deposition
may have to be taken into account. Although challenging, such modelling
would yield several advantages. First, it would increase our
understanding of the system, as the data would allow us to test
assumptions, and infer key biological properties as model parameters.
Second, it may yield a better-performing LR system, both because
extrapolations may be feasible and because a post-hoc calibration step
would no longer be needed. Third, it would increase our confidence in
the resulting LR-system, due to the biological interpretation of the
model.

The LR system might further be improved by a more sophisticated handling
of menstrual secretion. Biologically, menstrual secretion is a mixture
often containing blood and vaginal mucosa, as can be seen from the
similarity in markers that are expressed. However, in the current study
we ignored this biological knowledge. This could be improved by
explicitly modelling the relevant samples as a mixture that can contain
blood and vaginal mucosa as well as `atomic' menstrual secretion, which
triggers the expression of MMP10, MMP7 and MMP11. In case work, the
presence of vaginal mucosa or menstrual secretion often supports the
same scenario at activity level, while their absence may support the
other. Thus the approach we took here, combining the two body fluids
into one set of interest, is a practical way of working around the
special status of menstrual secretion. The relation between skin and
penile skin may be similar, with penile skin showing a similar pattern
of markers to skin but stronger expression of MUC4 (table 1): a marker
that is also used for the identification of vaginal mucosa.

Like all empirically based LR systems, the proposed system should not
yield very high or very low values of an LR, as these are based on
extrapolations that are not supported by the data. If we aim for being
maximally conservative, roughly speaking the highest LR supported is the
number of independent data points under H2 (see
\href{https://paperpile.com/c/A0Oh7K/ibeA}{{[}39{]}} for a more
sophisticated treatment). Furthermore, as we have seen the calculated
value of an LR for a given sample may fluctuate based on assumptions and
data used. We therefore recommend to only report the order of magnitude
of the LR, possibly using a verbal scale
\href{https://paperpile.com/c/A0Oh7K/oXFX}{{[}40{]}}, and to cap the LR
magnitude reported (as in figure 10). The cap depends on the values
supported by the size of the dataset, which in our case is in the range
of hundreds (max) and 1/hundreds (minimum).

\section{Conclusion}

We presented a way to construct LR systems for mRNA data that allow for
the presence of multiple body fluids, and found that a relatively simple
model based on logistic regression performs well, exhibits robustness
and is easily interpretable. We recommend using this system to calculate
weight of evidence from RNA data to support the forensic expert in case
work.

\section{Acknowledgements}

We would like to all volunteers for providing body fluid samples.

\section{Competing interests}

The authors have no competing interests to declare.

\newpage
\section{References}
\raggedright
\setlength{\parindent}{-0.35in}
\setlength{\leftskip}{0.35in}
\setlength{\parskip}{15pt}
\ 

{[}1{]} \href{http://paperpile.com/b/A0Oh7K/2Pxb}{K. Virkler, I.K.
Lednev, Analysis of body fluids for forensic purposes: from laboratory
testing to non-destructive rapid confirmatory identification at a crime
scene, Forensic Sci. Int. 188 (2009) 1--17.}

{[}2{]} \href{http://paperpile.com/b/A0Oh7K/sPWH}{T. Sijen, Molecular
approaches for forensic cell type identification: On mRNA, miRNA, DNA
methylation and microbial markers, Forensic Sci. Int. Genet. 18 (2015)
21--32.}

{[}3{]} \href{http://paperpile.com/b/A0Oh7K/gFjo}{A. Lindenbergh, M. de
Pagter, G. Ramdayal, M. Visser, D. Zubakov, M. Kayser, T. Sijen, A
multiplex (m)RNA-profiling system for the forensic identification of
body fluids and contact traces, Forensic Sci. Int. Genet. 6 (2012)
565--577.}

{[}4{]} \href{http://paperpile.com/b/A0Oh7K/7Fhk}{M. van den Berge, A.
Carracedo, I. Gomes, E.A.M. Graham, C. Haas, B. Hjort, P. Hoff-Olsen, O.
Maroñas, B. Mevåg, N. Morling, H. Niederstätter, W. Parson, P.M.
Schneider, D.S. Court, A. Vidaki, T. Sijen, A collaborative European
exercise on mRNA-based body fluid/skin typing and interpretation of DNA
and RNA results, Forensic Sci. Int. Genet. 10 (2014) 40--48.}

{[}5{]} \href{http://paperpile.com/b/A0Oh7K/ghVP}{C. Haas, B. Klesser,
C. Maake, W. Bär, A. Kratzer, mRNA profiling for body fluid
identification by reverse transcription endpoint PCR and realtime PCR,
Forensic Science International: Genetics. 3 (2009) 80--88.
https://doi.org/}\href{http://dx.doi.org/10.1016/j.fsigen.2008.11.003.}{10.1016/j.fsigen.2008.11.003.}

{[}6{]} \href{http://paperpile.com/b/A0Oh7K/4YSS}{J. Juusola, J.
Ballantyne, Multiplex mRNA profiling for the identification of body
fluids, Forensic Sci. Int. 152 (2005) 1--12.}

{[}7{]} \href{http://paperpile.com/b/A0Oh7K/bBbr}{A.D. Roeder, C. Haas,
mRNA profiling using a minimum of five mRNA markers per body fluid and a
novel scoring method for body fluid identification, Int. J. Legal Med.
127 (2013) 707--721.}

{[}8{]} \href{http://paperpile.com/b/A0Oh7K/z5VP}{B. Liu, Q. Yang, H.
Meng, C. Shao, J. Jiang, H. Xu, K. Sun, Y. Zhou, Y. Yao, Z. Zhou, H. Li,
Y. Shen, Z. Zhao, Q. Tang, J. Xie, Development of a multiplex system for
the identification of forensically relevant body fluids, Forensic Sci.
Int. Genet. 47 (2020) 102312.}

{[}9{]} \href{http://paperpile.com/b/A0Oh7K/sQry}{S. Ingold, G. Dørum,
E. Hanson, D. Ballard, A. Berti, K.B. Gettings, F. Giangasparo, M.-L.
Kampmann, F.-X. Laurent, N. Morling, W. Parson, C.R. Steffen, A. Ulus,
M. van den Berge, K.J. van der Gaag, V. Verdoliva, C. Xavier, J.
Ballantyne, C. Haas, Body fluid identification and assignment to donors
using a targeted mRNA massively parallel sequencing approach - results
of a second EUROFORGEN / EDNAP collaborative exercise, Forensic Sci.
Int. Genet. 45 (2020) 102208.}

{[}10{]} \href{http://paperpile.com/b/A0Oh7K/D9Sc}{R.I. Fleming, S.
Harbison, The development of a mRNA multiplex RT-PCR assay for the
definitive identification of body fluids, Forensic Sci. Int. Genet. 4
(2010) 244--256.}

{[}11{]} \href{http://paperpile.com/b/A0Oh7K/NRMs}{Y. Xu, J. Xie, Y.
Cao, H. Zhou, Y. Ping, L. Chen, L. Gu, W. Hu, G. Bi, J. Ge, X. Chen, Z.
Zhao, Development of highly sensitive and specific mRNA multiplex system
(XCYR1) for forensic human body fluids and tissues identification, PLoS
One. 9 (2014) e100123.}

{[}12{]} \href{http://paperpile.com/b/A0Oh7K/8Aqy}{M. van den Berge, B.
Bhoelai, J. Harteveld, A. Matai, T. Sijen, Advancing forensic RNA
typing: On non-target secretions, a nasal mucosa marker, a differential
co-extraction protocol and the sensitivity of DNA and RNA profiling,
Forensic Sci. Int. Genet. 20 (2016) 119--129.}

{[}13{]} \href{http://paperpile.com/b/A0Oh7K/zYtd}{A. Lindenbergh, M.
van den Berge, R.-J. Oostra, C. Cleypool, A. Bruggink, A. Kloosterman,
T. Sijen, Development of a mRNA profiling multiplex for the inference of
organ tissues, Int. J. Legal Med. 127 (2013) 891--900.}

{[}14{]} \href{http://paperpile.com/b/A0Oh7K/aV2v}{J. de Zoete, J.
Curran, M. Sjerps, Categorical methods for the interpretation of RNA
profiles as cell type evidence and their limitations, Forensic Science
International: Genetics Supplement Series. 5 (2015) e305--e307.}

{[}15{]} \href{http://paperpile.com/b/A0Oh7K/w35J}{J. de Zoete, J.
Curran, M. Sjerps, A probabilistic approach for the interpretation of
RNA profiles as cell type evidence, Forensic Sci. Int. Genet. 20 (2016)
30--44.}

{[}16{]} \href{http://paperpile.com/b/A0Oh7K/v1GQ}{G. Dørum, S. Ingold,
E. Hanson, J. Ballantyne, G. Russo, S. Aluri, L. Snipen, C. Haas,
Predicting the origin of stains from whole miRNome massively parallel
sequencing data, Forensic Sci. Int. Genet. 40 (2019) 131--139.}

{[}17{]} \href{http://paperpile.com/b/A0Oh7K/aQCk}{G. Dørum, S. Ingold,
E. Hanson, J. Ballantyne, L. Snipen, C. Haas, Predicting the origin of
stains from next generation sequencing mRNA data, Forensic Sci. Int.
Genet. 34 (2018) 37--48.}

{[}18{]} \href{http://paperpile.com/b/A0Oh7K/Y94R}{D. Iacob, A. Fürst,
T. Hadrys, A machine learning model to predict the origin of
forensically relevant body fluids, Forensic Science International:
Genetics Supplement Series. 7 (2019) 392--394.}

{[}19{]} \href{http://paperpile.com/b/A0Oh7K/I25b}{S. Fujimoto, S.
Manabe, C. Morimoto, M. Ozeki, Y. Hamano, E. Hirai, H. Kotani, K.
Tamaki, Distinct spectrum of microRNA expression in forensically
relevant body fluids and probabilistic discriminant approach, Sci. Rep.
9 (2019) 14332.}

{[}20{]} \href{http://paperpile.com/b/A0Oh7K/Gv5Y}{J.
Gonzalez-Rodriguez, P. Rose, D. Ramos, D.T. Toledano, J. Ortega-Garcia,
Emulating DNA: Rigorous Quantification of Evidential Weight in
Transparent and Testable Forensic Speaker Recognition, IEEE Trans. Audio
Speech Lang. Processing. 15 (2007) 2104--2115.}

{[}21{]} \href{http://paperpile.com/b/A0Oh7K/X3hh}{A. van Es, W. Wiarda,
M. Hordijk, I. Alberink, P. Vergeer, Implementation and assessment of a
likelihood ratio approach for the evaluation of LA-ICP-MS evidence in
forensic glass analysis, Sci. Justice. 57 (2017) 181--192.}

{[}22{]} \href{http://paperpile.com/b/A0Oh7K/zukR}{W. Bosma, S. Dalm, E.
van Eijk, R. El Harchaoui, E. Rijgersberg, H.T. Tops, A. Veenstra, R.
Ypma, Establishing phone-pair co-usage by comparing mobility patterns,
Sci. Justice. 60 (2020) 180--190.}

{[}23{]} \href{http://paperpile.com/b/A0Oh7K/Ijw7}{G. Tsoumakas, I.
Katakis, I. Vlahavas, Mining Multi-label Data, in: O. Maimon, L. Rokach
(Eds.), Data Mining and Knowledge Discovery Handbook, Springer US,
Boston, MA, 2010: pp. 667--685.}

{[}24{]} \href{http://paperpile.com/b/A0Oh7K/6g1L}{T. Hastie, R.
Tibshirani, J. Friedman, The Elements of Statistical Learning: Data
Mining, Inference, and Prediction, Second Edition, Springer Science \&
Business Media, 2009.}

{[}25{]} \href{http://paperpile.com/b/A0Oh7K/ZSpg}{B. Scholkopf, A.J.
Smola, Learning with Kernels: Support Vector Machines, Regularization,
Optimization, and Beyond, MIT Press, Cambridge, MA, USA, 2001.}

{[}26{]} \href{http://paperpile.com/b/A0Oh7K/gaqz}{J.H. Friedman, Greedy
Function Approximation: A Gradient Boosting Machine, Ann. Stat. 29
(2001) 1189--1232.}

{[}27{]} \href{http://paperpile.com/b/A0Oh7K/8zdu}{T. Chen, C. Guestrin,
Xgboost: A scalable tree boosting system, in: Proceedings of the 22nd
Acm Sigkdd International Conference on Knowledge Discovery and Data
Mining, 2016: pp. 785--794.}

{[}28{]} \href{http://paperpile.com/b/A0Oh7K/Pj4Z}{Tin Kam Ho, Random
decision forests, in: Proceedings of 3rd International Conference on
Document Analysis and Recognition, 1995: pp. 278--282 vol.1.}

{[}29{]} \href{http://paperpile.com/b/A0Oh7K/USmn}{F. Pedregosa, G.
Varoquaux, A. Gramfort, V. Michel, B. Thirion, O. Grisel, M. Blondel, P.
Prettenhofer, R. Weiss, V. Dubourg, J. Vanderplas, A. Passos, D.
Cournapeau, M. Brucher, M. Perrot, E. Duchesnay, Scikit-learn: Machine
Learning in Python, J. Mach. Learn. Res. 12 (2011) 2825--2830.}

{[}30{]} \href{http://paperpile.com/b/A0Oh7K/BHRM}{J. Platt,
Probabilistic outputs for support vector machines and comparisons to
regularized likelihood methods, in: A.J. Smola, P. Bartlett, B.
Scholkopf, D. Schuurmans (Eds.), Advances in Large Margin Classifiers,
MIT Press, Cambridge, 1999: pp. 61--74.}

{[}31{]} \href{http://paperpile.com/b/A0Oh7K/LdyY}{A.P. Dawid, The
Well-Calibrated Bayesian, Journal of the American Statistical
Association. 77 (1982) 605--610.
https://doi.org/}\href{http://dx.doi.org/10.1080/01621459.1982.10477856.}{10.1080/01621459.1982.10477856.}

{[}32{]} \href{http://paperpile.com/b/A0Oh7K/GnLs}{D. Ramos, J.
Gonzalez-Rodriguez, Reliable support: Measuring calibration of
likelihood ratios, Forensic Science International. 230 (2013) 156--169.
https://doi.org/}\href{http://dx.doi.org/10.1016/j.forsciint.2013.04.014.}{10.1016/j.forsciint.2013.04.014.}

{[}33{]} \href{http://paperpile.com/b/A0Oh7K/JLBY}{P. Vergeer, I.
Alberink, M. Sjerps, R. Ypma, Why calibrating LR-systems is best
practice. A reaction to ``The evaluation of evidence for
microspectrophotometry data using functional data analysis'', in FSI
305, Forensic Sci. Int. 314 (2020) 110388.}

{[}34{]} \href{http://paperpile.com/b/A0Oh7K/7cqN}{D.D. Lewis, W.A.
Gale, A Sequential Algorithm for Training Text Classifiers, arXiv
{[}cmp-Lg{]}. (1994).}
\url{http://arxiv.org/abs/cmp-lg/9407020}\href{http://paperpile.com/b/A0Oh7K/7cqN}{.}

{[}35{]} \href{http://paperpile.com/b/A0Oh7K/aEuf}{G.S. Morrison,
Tutorial on logistic-regression calibration and fusion:converting a
score to a likelihood ratio, Australian Journal of Forensic Sciences. 45
(2013) 173--197.
https://doi.org/}\href{http://dx.doi.org/10.1080/00450618.2012.733025.}{10.1080/00450618.2012.733025.}

{[}36{]} \href{http://paperpile.com/b/A0Oh7K/wFZV}{D. Meuwly, D. Ramos,
R. Haraksim, A guideline for the validation of likelihood ratio methods
used for forensic evidence evaluation, Forensic Sci. Int. 276 (2017)
142--153.}

{[}37{]} \href{http://paperpile.com/b/A0Oh7K/3T03}{N. Brümmer, J. du
Preez, Application-independent evaluation of speaker detection, Comput.
Speech Lang. 20 (2006) 230--275.}

{[}38{]} \href{http://paperpile.com/b/A0Oh7K/P2lO}{J. Read, B.
Pfahringer, G. Holmes, E. Frank, Classifier chains for multi-label
classification, Mach. Learn. 85 (2011) 333.}

{[}39{]} \href{http://paperpile.com/b/A0Oh7K/ibeA}{P. Vergeer, A. van
Es, A. de Jongh, I. Alberink, R. Stoel, Numerical likelihood ratios
outputted by LR systems are often based on extrapolation: When to stop
extrapolating?, Sci. Justice. 56 (2016) 482--491.}

{[}40{]} \href{http://paperpile.com/b/A0Oh7K/oXFX}{ENFSI (European
network of forensic science institutes), ENFSI guideline for evaluative
reporting in forensic science, (2015).}
\url{http://enfsi.eu/wp-content/uploads/2016/09/m1_guideline.pdf}\href{http://paperpile.com/b/A0Oh7K/oXFX}{.}

\end{document}